\documentstyle[preprint,prl,aps]{revtex}
\begin{document}

\title{Relaxation and Landau-Zener experiments down to 100 mK in Ferritin}
\author{M. Dur\'{a}n, E. del Barco, J. M. Hern\'{a}ndez and J. Tejada}
\address{Departamento de F\'\i sica Fundamental, Universitat de Barcelona\\
Diagonal 647, Barcelona, 08028, Spain\\}
\date{15/9/01}
\maketitle

\begin{abstract}
Temperature-independent magnetic viscosity in ferritin has been
observed from $\sim$ 2 K down to 100 mK, proving that quantum
tunneling plays the main role in these particles at low
temperature. Magnetic relaxation has also been studied using the
Landau-Zener method making the system crossing zero resonant field
at different rates, $\alpha=dH/dt$, ranging from $10^{-5}$ to
$10^{-3}$ T/s, and at different temperatures, from 150 mK up to
the blocking temperature. We propose a new $Tln(\Delta
H_{eff}/\tau_0\alpha)$ scaling law for the Landau-Zener
probability in a system distributed in volumes, where $\Delta
H_{eff}$ is the effective width of the zero field resonance.
\end{abstract}


Over the past decade there have been experimentally observed a
large number of quantum phenomena in the dynamics of the magnetic
moment of mesoscopic systems. Monodomain magnetic particles of
nanometer size contain thousands of magnetic atoms strongly
interconnected by exchange interaction. As a result of the
exchange interaction, all the atomic spins align parallel or
antiparallel between them, resulting in a ferro, or
ferri-antiferromagnetic ordering respectively. One particular
system has had much attention since Awschalom and co-workers
announced the observation of a resonance near 1 MHz that
interpreted in terms of quantum coherence of the magnetic moment
\cite{Awschalom,Gider}: the system is composed by
antiferromagnetic particles which grow inside the cage of the
horse spleen ferritin proteins \cite{Gider2}.

Next experimental studies of the dynamics of the magnetization of
ferritin particles, carried out at kelvin regime, showed different
phenomena interpreted as quantum tunneling of the ferritin
magnetic moment
\cite{Gider2,Tejada1,Friedman,Tejada2,Salah,delBarco,Luis,Sappey}.
These phenomena can be differentiated as follows: The first was
the temperature-independent relaxation of the magnetization below
2.3 K observed in the magnetic viscosity measurements
\cite{Tejada1}. Secondly, a non-monotonic behaviour of the
blocking temperature, $T_B$, on the magnetic field
\cite{Gider,Friedman,Tejada2,delBarco,Luis,Sappey}, contrasting
with the monotonic classical expectation. Thirdly, a maximum at
zero field in the magnetic field derivative of the magnetization
extracted from the hysteresis cycles over 4 K
\cite{Gider,Friedman,Tejada2,Salah,delBarco}. And finally, a rapid
increase of the magnetic viscosity as the magnetic field
approaches to zero at temperatures higher than 3 K
\cite{Tejada2,delBarco,Luis}. More recently, there have been done
$^{57}$Fe M\"{o}ssbauer spectroscopy measurements on an artificial
ferritin sample down to 50 mK showing non-incoherent tunnel
fluctuations around $10^8$ Hz \cite{Bonville}. Attending to the
recent observation of resonant quantum tunneling of the spin
observed in molecular clusters like Mn$_{12}$-Acetate
\cite{Friedman2,Hernandez}, one can conclude that all these
phenomena observed in ferritin particles in the kelvin range can
be attributed to thermally activated resonant quantum tunneling of
the magnetic moment at zero field. The observation of tunneling
only at zero field lies in the fact that ferritin has a broad
distribution of energy barriers due to the distribution of volumes
of the particles ($f(U=KV)$, where $K$ is the anisotropy
constant), and their anisotropy axes are randomly oriented. In the
other hand, the temperature-independent viscosity below the
crossover temperature, $T_c$ = 2.3 K, indicates that, below this
temperature, the quantum magnetic relaxation of ferritin occurs
through the lowest states.

In this paper we present new magnetic data which extend the
quantum relaxation measurements to the millikelvin regime. At the
same time, in order to estimate the value of the quantum
splittings of ferritin particles, we have done measurements of the
change of magnetization when the system crosses zero magnetic
field at different rates of the field sweep and we analyze the
results in terms of the Landau-Zener probability associated to the
magnitude of the splitting playing the main role in the quantum
relaxation. This is the same method used by Wernsdorfer et al. to
determine the quantum splittings of Fe$_8$ molecular clusters
\cite{Wernsdorfer}.

Ferritin is an iron storage protein. It has a spherical cage of
about 8 nm in diameter in whose interior grows mineral
ferrihydrite combined with a phosphate. Its core is equivalent to
a small antiferromagnetic particle. The size of the core in
natural ferritin ranges from 3 to 7.5 nm. The fully packed
ferritin contains 4500 $Fe^{3+}$ ions. A small magnetic moment of
the particle arises from the non-compensation of collinear spin
sublattices due to the finite size and irregular shape of the
core. The spin of the sublattice, $S$, is of the order of 5000,
while the non-compensated spin, $s$, is below 100. This number
corresponds to 15 non-compensated $Fe^{3+}$ ions extracted from
magnetic susceptibility measurements \cite{Tejada1}, in agreement
with the theoretical expectation for the square root of the number
of ions at the surface of the particle, $N_S \sim(4500)^{2/3}$,
which gives $(4500)^{1/3} \sim$ 16. This non-compensated spin
looks in one of two directions along the anisotropy axis of the
particle. In our experiments we have used a {\it Fluka
Biochemical} diluted natural ferritin sample. The distribution of
volumes of the sample, $f(V)dV$, is plotted in the inset of figure
1 (extracted from reference \cite{Luis}). The center volume is
$V_0 \sim$ 150 nm$^3$. The low moment of the antiferromagnetic
particles makes the interactions between different particles to be
negligible, as one can see in the inverse susceptibility in the
superparamagnetic regime that extrapolates to zero at
$T{\rightarrow }0$ \cite{Tejada2}.

Low temperature magnetic relaxation measurements were done in an
{\it Oxford Instruments} $^3$He-$^4$He dilution cryostat in the
following manner: The sample is cooled until the measure
temperature and then a magnetic field of 1 T is applied during 10
minutes. After that, the field is switched off and the
magnetization is measured during 3 hours. In order to avoid
remanent fields in the superconducting magnet and to obtain the
relaxation measurements as close as possible to zero field a
demagnetizing cycle is immediately applied after switching off the
field. The demagnetizing cycle was previously tested in a pure Pb
diamagnetic sample. This method makes the field along the major
part of the relaxation to be zero with a precision of $\pm 1$ Oe.
The measurements were done at different temperatures ranging from
100 mK up to 1,5 K in the dilution cryostat and repeated in the
same manner in a {\it Quantum Design MPMS} magnetometer at
temperatures up to 25 K. The logarithmic on time dependence of the
magnetic relaxation is clearly observed over the whole measure. In
a sample with a distribution of energy barriers, the quantitative
magnitude which measures the relaxation time is the magnetic
viscosity defined as \cite{Chudnovsky}
\begin{equation}
S=\frac 1{(M_{ini}(H,T)-M_{eq}(H,T))}\frac{dM(H,T,t)}{d{\ln
}(t)}\;\;\;,
\end{equation}
where $M_{eq}(H,T)$ is the equilbrium magnetization of the system
at fixed temperature and field, which is $M_{eq}(H,T)$ = 0 in our
case, and $M_{ini}(H,T)$ is the initial magnetization. In our
experiments $M_{ini}(H,T)$ was taken from the extrapolation at
small time of each magnetic relaxation curve. It is known that
after switching off the field the system rapidly runs to a
critical state in a time much more shorter than the times involved
in the slow relaxation process occurring after the system reaches
this critical state and relaxes to the final equilibrium state
\cite{Chudnovsky}. The observed dependence of the magnetic
viscosity with temperature is shown in figure 1. The viscosity
shows a maximum at $T_B$ $\sim$ 10 K. This is the blocking
temperature, defined as
\begin{equation}
T_B=\frac {KV}{{\ln }({t_m}/{\tau }_0)}\;\;\;,
\end{equation}
which for viscosity measurements, with a characteristic measuring
time, $t_m$, of hours, corresponds to the unfreezing of the
magnetic moment of a particle of volume $V$ which changes its
orientation jumping over the energy barrier. In a sample
distributed in size (see inset of figure 1) there is a
distribution of energy barriers, $f(U=KV)$. The rate at which
individual moments of the particles jump across the anisotropy
barrier depends on temperature through the Arrhenius exponential
factor, $exp(-KV/K_BT)$. The maximum observed in the viscosity at
$T_B$ corresponds to the unfreezing of the particles having the
center volume, $V_0$, of the volume distribution. If we look at
the volume distribution of figure 1, the blocking temperature may
correspond to the particles with a volume around 150 nm$^3$. For
these particles, using eq. (2), with $\tau_0 \sim 10^{-8}$ s,
$t_m$ = 10$^4$ s and $K = 2.5\times10^{-5}$ erg/cm$^3$, we obtain
$T_B \sim$ 10 K, in good agreement with the experimental result.
As the temperature decreases, the magnetic viscosity goes to zero,
as expected for thermal relaxation in a system with barriers
distribution. However, below $\sim$ 2 K the viscosity becomes
independent on temperature down to 100 mK. This temperature, at
which the system crosses from thermal to quantum relaxation regime
is called crossover temperature, $T_c$ \cite{Tejada1}. The new
data showed in this paper extend the observation of the plateau of
the magnetic viscosity down to a few millikelvin. This takes high
relevance assuming the fact that below $T_c$ the system relaxes
exclusively through the lowest levels of the magnetic structure by
quantum tunneling. This temperature does not depend on the volume
of the particles. The expression expected from theory which
determines this temperature for antiferromagnetic monodomain
particles is $T_c \sim (2\epsilon_{an}\epsilon_{ex})^{1/2}/2\pi$
\cite{Barbara}. Taking $\epsilon_{an} \sim $ 0.1 K (anisotropy
energy per spin) and $\epsilon_{ex} \sim 10^3$ K (exchange energy
per atom) \cite{Tejada2}, we obtain $T_c \sim$ 2 K in good
agreement with the experimental value.

ZFC magnetization measurements have been done from 100 mK up to 25
K. The measurements were done as follows: First the system is
cooled from room temperature down to 100 mK in the absence of
magnetic field. Due to the random orientation of the anisotropy
axes of the particles the total magnetization is zero. Then a
small magnetic field, $H$ = 200 Oe, is applied and the
magnetization is measured as the temperature is increased. The
result is shown in figure 2. In this figure we can observe the
maximum corresponding to the blocking temperature of the sample,
$T_B \sim$ 13 K, in agreement with eq. (2) using $t_m$ = 10 s. A
significant fact that one can observe in this figure is the
increase of the magnetization as the temperature goes to zero
below $\sim 0.5$ K. The behaviour of the magnetization in this
temperature range is $1/T$, indicating that there is a
superparamagnetic contribution to the total magnetization. To
understand this let us see the expected behaviour for a sample
with a broad distribution of barriers in the thermal regime. After
the field is applied at the lowest temperature, the smallest
particles, having the smallest barriers, become free to rotate
their magnetic moments out of the anisotropy axis and align them
with respect to the applied field, contributing to the net
magnetization of the sample. As the temperature increases, bigger
and bigger particles can jump across the anisotropy barrier and
the total magnetization increases. When the major part of the
magnetic moments of the particles are free, over $T_B$, the system
behaves paramagnetically following the $1/T$ Curie behaviour. This
is opposed to the observed decrease of $M(T)$ at low temperature.
Two different facts can explain this paradox: (a) thermal
superparamagnetic behaviour of a second distribution of particles
of smaller size or (b) quantum superparamagnetic behaviour of the
smallest particles of the mono-distributed sample. The first
explanation does not agree with the distribution showed in figure
1, where there is not a significant number of particles below 50
nm$^3$. Applying eq. (2) we obtain that a significant number of
particles smaller than 5 nm$^3$ is needed to obtain $T_B < $ 0.1
K. The ac-susceptibility measurements (window time of 10$^{-3}$ s)
\cite{Luis} and M\"{o}ssbauer espectroscopy (10$^{-8}$ s)
\cite{Bonville,Kilcoyne} show that there are not significant
particles of this size behaving superparamagnetically at low
temperature. The second explanation, quantum superparamagnetism,
explains better the $1/T$ behaviour at low temperature. This is
the quantum behaviour of the particles for which $T_B$ is smaller
than $T_c$. That is, these particles do not feel the anisotropy
barrier because they can rotate their magnetic moments by quantum
tunneling even if the temperature is not enough to jump across the
barrier. The quantum tunneling rate is determined by the WKB
exponent, $B \sim KV/K_BT_c$, in the following manner, $\Gamma
\sim exp(-B)$. This means that the particles having smallest size
have the higher probability to tunnel across the barrier.

The most direct way to measure the quantum tunneling splitting,
$\Delta$, is by using the Landau-Zener model \cite{Zener}, which
gives the tunnel probability, $P$, when a resonance is crossed at
a given sweeping rate, $\alpha$:
\begin{equation}
P =1-exp\left[-\frac {h\Delta^2}{2g\mu_BS\alpha}\right]\;\;\;,
\end{equation}
where $h$ is the Plank's constant and $S$ is the spin of a
particle. Due to the distribution of volumes in ferritin there are
a distribution of spin values, $S(V)$, and a distribution of
quantum splittings, $\Delta(V)$. Also, the random orientation of
the anisotropy axis of the particles in the sample introduces a
distribution of sweeping rates, $\alpha(\theta)$, on the angle
between the applied field and the anisotropy axis of each
particle. This makes that different particles have different
tunnel probability at a given sweeping rate depending in both
volume and orientation respect to the applied magnetic field.
Taking into account the mentioned conditions, we can express the
change of magnetization of the whole sample as the zero resonant
field is crossed from $H_i$ to $H_f$ at a given $\alpha$ in terms
of the Landau-Zener probability as follows:
\begin{equation}
\frac{M_f-M_i}{M_{eq}-M_i}=2\pi \int_0^{\pi/2}
sin(\theta)d{\theta}\int_V
S[V]P[\Delta(V),S(V),\alpha(\theta)]f[V]dV\;\;\;,
\end{equation}
where $M_i$, $M_f$ and $M_{eq}$ are the initial, final and
equilibrium magnetizations, respectively. The integral over
$\theta$ has been chosen to take into account the random
orientation of the anisotropy axes of the particles respect to the
applied field. The form of $\alpha(\theta)$ for one particle is
then $cos(\theta)\alpha$.

Our experiments where done in the following manner: First, a
saturating magnetic field was applied at the measure temperature.
Then, the field was changed to $H_i$ = 250 Oe at the highest
sweeping rate and the magnetization was measured giving $M_i$.
Immediately, the field was changed to $H_f$ = -250 Oe at a given
$\alpha$, measuring $M_f$ after the process was finished. The
procedure was repeated at different sweeping rates, ranging from
$10^{-5}$ T/s up to $10^{-3}$ T/s and at different temperatures,
from 100 mK up to the blocking temperature. The results are shown
in figure 3. In order to make the nomenclature shorter we will use
$P_{\Delta M}$ (probability to change the magnetization) instead
the expression given in eq. (4). One can see that, at a given
temperature, $P_{\Delta M}$ increases when $\alpha$ decreases.
That is, as the zero field resonance is crossed slower the
probability to change the magnetization of the sample is higher.
With the same dependence in $\alpha$, the probability becomes
higher for higher temperatures. The bahaviour of $P_{\Delta M}$ on
1/$\alpha$ is perfectly logarithmic. This dependence reminds the
behaviour of the time magnetic relaxation observed in this sample
and, in general, in any sample with barriers distribution. Indeed,
we can find the equivalence between the sweeping rate and time
using $t=\Delta H/\alpha$, where $\Delta H = H_i-H_f$ = 500 Oe in
our experiment. Due to this equivalence we can define a new
parameter, $S_{LZ}$, to evaluate the characteristics of the
magnetic relaxation in a Landau-Zener process with a barrier
distributed sample, in the same manner that the magnetic viscosity
does it in time magnetic relaxations. That is, $S_{LZ}$ can be
expressed as
\begin{equation}
S_{LZ}=\frac {dP_{\Delta M}}{d\ln(\Delta H/\alpha)}\;\;\;,
\end{equation}
where $P_{\Delta M}=(M_f-M_i)/(M_{eq}-M_i)$. The temperature
dependence of the {\it Landau-Zener viscosity}, $S_{LZ}$, is shown
in the inset of figure 4. From the comparison of this result with
the magnetic viscosity extracted from time magnetic relaxations
(figure 1), the agreement between the results of both methods is
clearly observed. $S_{LZ}$ has a maximum at 10 K, which, using eq.
(2), corresponds to the blocking temperature with an effective
time, $\Delta H/\alpha$, of 10$^4$ s, this is, $\alpha$ =
10$^{-5}$ T/s and $\Delta H$ = 5$\times$10$^{-2}$ T. That is, the
Landau-Zener procedure carried out in a sample of particles
distributed in volume gives the same information that the magnetic
viscosity analysis. However, we can extract new information from
this procedure if we analyze the change of magnetization as the
zero field resonance is crossed under a new scaling law proposed
following.

To look for evidences of thermal or quantum relaxation from time
magnetic relaxation experiments it is usually used a
$T\ln(t/\tau_0)$ plot. In the thermal relaxation regime the
dependence of the magnetization on $T\ln(t/\tau_0)$ scales in a
master curve if the characteristic relaxation time, $\tau_0$, is
adequately chosen. This analysis permits to extract evidences of
quantum tunneling when the magnetic relaxation departs from the
master curve below a characteristic temperature. This temperature
corresponds to the crossover temperature, $T_c$, explained before.
Independently, it is possible to find the barriers distribution
function from the derivative of the master curve. We show in
figure 4 (black lines and right axis) the $T\ln(t/\tau_0)$ plot
corresponding to the time magnetic relaxations of the ferritin
sample. The scaling is reached using $\tau_0$ = 10$^{-8}$ s. One
can see that the relaxation curves depart from the master curve at
temperatures below 5 K, indicating the presence of quantum
tunneling as the temperature arrives near the crossover
temperature. Below $\sim$2 K all the curves are parallel, showing
the temperature independence of the quantum relaxation.

We propose a new scaling law, equivalent to the $T\ln(t/\tau_0)$
plot in time magnetic relaxations, for the total change of
magnetization of a volume distributed sample in a Landau-Zener
process using a $T\ln(t_{eff}/\tau_0)$ plot, where $t_{eff}$ =
$\Delta H_{eff}/\alpha$. In figure 4 (open circles and left axis)
is shown the $T\ln(t_{eff}/\tau_0)$ plot of $P_{\Delta
M}^{not}=1-P_{\Delta M}$, in order to compare with the magnetic
relaxation master curve. The scaling is obtained with $\tau_0$ =
10$^{-8}$ s and $\Delta H_{eff}$ = 5 Oe. It is observed that the
data collapse into a master curve for temperatures higher than
$\sim$5 K. The value of the effective resonance width, $\Delta
H_{eff}$ = 5 Oe, is two orders of magnitude smaller than the width
of the zero resonance observed in the magnetic hysteresis loops at
the same temperatures, $\Delta H\sim$ 1000 Oe
\cite{Gider,Friedman,Tejada2,Salah,delBarco} and associated to
thermally assisted resonant quantum tunneling
\cite{Tejada2,delBarco,Luis}. The same phenomena was previously
observed in molecular clusters \cite{Friedman2,Hernandez}. In
principle, the width  of this resonance is associated to the
quantum splitting of the blocking level, $m_B$. This is the level
through which the quantum tunneling occurs at a given temperature.
In ferritin the width of the resonance is associated to the
distribution of quantum splittings of the blocking levels due to
the different volumes of the particles of the sample. This fact,
together with the random orientation of the anisotropy axes
respect to the applied magnetic field, makes the width of the zero
field resonance to be several orders of magnitude higher than the
width of the quantum splitting of one of the particles of the
sample. However, the scaling law proposed here takes into account
the effect of an average particle of the sample. Due to this, the
physical meaning of $\Delta H_{eff}$ extracted from the master
curve can be attributed to the width of the zero field resonance
for an average particle of the sample. That is, we may associate
$\Delta H_{eff}$ with the quantum splitting of the effective
blocking level, $\Delta_{eff}$, of the distribution of particles
in ferritin in the following manner: $\Delta_{eff}\sim
g\mu_BS\Delta H_{eff}$. Using $S\sim$ 50, we obtain
$\Delta_{eff}\sim$ 700 MHz. Taking into account the uncertainties
associated to the random orientation of the anisotropy axes of the
particles it seems clear that the obtained value of the quantum
splitting of the effective blocking level agrees with the $\sim$1
MHz resonance found by Awschalom et al. \cite{Awschalom} and
attributed to the quantum splitting of the ground state of
ferritin particles.

In conclusion, we have obtained temperature-independent magnetic
relaxation from both magnetic viscosity measurements and {\it
Landau-Zener viscosity} down to 100 mK. We have proposed a new
scaling law for the probability to change the magnetization in a
Landau-Zener process in a sample with barriers distribution. The
excellent agreement between the two studied methods (time magnetic
relaxation and Landau-Zener process) permits to establish that
quantum tunneling is the process governing magnetic relaxation in
ferritin at low temperatures. From the comparison of these two
methods one can extract additional information about the magnitude
of the effective quantum splitting playing the main role in the
low-temperature magnetic relaxation of a sample distributed in
volume.

The authors thank the financial support from IST-1999-29110 and
MEC Grant number PB96-0169.

\pagebreak

{\bf FIGURE CAPTIONS}\\

{\bf Figure 1:} Magnetic viscosity as a function of temperature of ferritin sample
extracted from time magnetic relaxations at zero field. The inset shows the
distribution of volumes of the sample (extracted from \cite{Luis}).\\

{\bf Figure 2:} ZFC-FC magnetization versus temperature in
ferritin recorded at $H$ = 200 Oe. The inset amplifies the low
temperature zone. 1/$T$ quantum superparamagnetism behaviour is
observed below 0.5 K.\\

{\bf Figure 3:} Probability to change the magnetization as the zero field resonance
is crossed at different sweeping rates of the applied magnetic field and at different
temperatures. The dependence with 1/$\alpha$ is clearly logarithmic.\\

{\bf Figure 4:} $T\ln(t_{eff}/\tau_0)$ plot for both time magnetic
relaxations, $t_{eff}=t$ (black lines, right axis), and
Landau-Zener relaxations, $t_{eff}=\Delta H_{eff}/\alpha$ (open
circles, left axis). The values of $\tau_0$ and $\Delta H_{eff}$
used to obtain the scaling are $10^{-8}$ s and 5 Oe, respectively.
The inset shows the temperature dependence of the {\it
Landau-Zener viscosity}.\\

\pagebreak

\end{document}